\documentclass[%
preprint,
superscriptaddress,
 amsmath,amssymb,
 aps,
prl,
floatfix,
]{revtex4-1}

\usepackage{pdfpages}
\usepackage{graphicx}% Include figure files
\usepackage{dcolumn}% Align table columns on decimal point
\usepackage{bm}% bold math
\usepackage{verbatim}
\usepackage{color,soul}
\usepackage{epstopdf}
\AppendGraphicsExtensions{.tif}

\makeatletter
\AtBeginDocument{\let\LS@rot\@undefined}
\makeatother
\begin{document}

%\preprint{APS/123-QED}

\title{Superconducting Phase of $\mathrm{Ti_xO_y}$ Thin Films Grown by Molecular Beam Epitaxy }% Force line breaks with \\
\author{Yasemin Ozbek}
 \affiliation{Department of Physics, North Carolina State University, Raleigh, NC, 27695, USA}

\author{Cooper Brooks}
 \affiliation{Department of Physics, North Carolina State University, Raleigh, NC, 27695, USA}

\author{Xuanyi Zhang}
 \affiliation{Department of Physics, North Carolina State University, Raleigh, NC, 27695, USA}
 
\author{Athby Al-Tawhid}
 \affiliation{Department of Physics, North Carolina State University, Raleigh, NC, 27695, USA}

\author{Vladmir A. Stoica}
\affiliation{Department of Materials Science and Engineering and Materials Research Institute, Pennsylvania State University, University Park, PA 16802, USA}
 
\author{Zhan Zhang}
\affiliation{Advanced Photon Source, Lemont, IL 76019, USA}

\author{Divine P. Kumah}
\email{dpkumah@ncsu.edu}
 \affiliation{Department of Physics, North Carolina State University, Raleigh, NC, 27695, USA}%Lines break automatically or can be forced with \\
%\affiliation{%

\date{\today}% It is always \today, today,

\begin{abstract}
We investigate the complex relationship between the growth conditions and the structural and transport properties of $\mathrm{Ti_xO_y}$ thin films grown by molecular beam epitaxy. Transport properties ranging from metallicity to superconductivity and insulating states are stabilized by effectively tuning the O/Ti ratio via the Ti flux rate and the O partial pressure, $P_{Ox}$, for films grown on (0001)-$\mathrm{Al_2 O_3}$ substrates at 850$^{\circ}$ C. A cubic $c-\mathrm{TiO_{1\pm\delta}}$ buffer layer is formed for low O/Ti ratios while a corundum cr-$\mathrm{Ti_2 O_3}$ layer is formed under higher oxidizing conditions. Metallicity is observed for c-$\mathrm{TiO_{1-\delta}}$ buffer layers. The superconducting $\mathrm{\gamma-Ti_3 O_5}$ Magn\'eli phase is found to nucleate on a c-$\mathrm{TiO_{1-\delta}}$ buffer for intermediate $P_{Ox}$ conditions and an insulator-superconducting transition is observed at 4.5 K (T$_C^{onset}=6 K$) for 85 nm thick films. Strain relaxation of the $\mathrm{\gamma-Ti_3 O_5}$ occurs with increasing film thickness and correlates with a thickness-dependent increase in T$_C$ observed for $\mathrm{Ti_xO_y}$ thin films. %A mixed-phase film is formed for O/Ti ratios close to the c-TiO/$\mathrm{Ti_2 O_3}$ phase boundary.  These results are critical for understanding the intimate link between growth conditions and film thickness on the superconducting properties of Ti$_x$O$_y$ thin films.

%\note{how do we connect this to the main point of the paper?}

\end{abstract}

\maketitle

\section{Introduction}

The simple binary oxide, titanium oxide, Ti$_x$O$_y$, forms a wide range of polymorphs and Magn\'eli (Ti$_n$O$_{n-1}$) phases with properties ranging from insulating rutile and anatase TiO$_2$, to corundum $\mathrm{cr-Ti_2 O_3}$ which undergoes a metal-insulator transition at 450 K, to superconducting cubic NaCl-type c-TiO (bulk T$_C=$ 1-2 K).\cite{ doyle1968vacancies,iwasaki1969polymorphism, hulm1972superconductivity, mclachlan1982new, reed1972superconductivity} The ability to utilize the epitaxial stabilization of a given phase in thin films synthesised with atomic-layer control provides a route to tune the electronic properties of Ti$_x$O$_y$.\cite{zhang2017enhanced, wagner1997growth, li2021single, alexander1990tiox, li2019electronic, li2018orthorhombic, yoshimatsu2020metallic} Recent reports of superconductivity in Ti$_x$O$_y$ thin films as high as 11 K has sparked renewed interest in superconducting binary oxides.\cite{xu2018nano} The transition temperature, $T_C$ has been reported to depend on the film thickness,\cite{zhang2019quantum, li2018observation} oxygen stoichiometry\cite{fan2018quantum, fan2019structure} and growth temperature\cite{li2018observation}. The enhanced T$_C$ in thin films was attributed to the formation of a high growth-temperature orthorhombic $\mathrm{o-Ti_2O_3}$ phase\cite{li2018observation} (T$_C=$ 8 K), excess oxygen in c-$\mathrm{TiO_{1+\delta}}$ films (T$_C=$ 7.4 K)\cite{zhang2017enhanced, fan2018quantum}, interfacial superconductivity at stoichiometric c-TiO/non-stoichiometric $\mathrm{TiO_{1+\delta}}$ interfaces in core-shell structures (T$_C=$ 11 K)\cite{xu2018nano}, and the epitaxial stabilization of the $\mathrm{\gamma-Ti_3 O_5}$ (T$_C=$ 7 K)\cite{yoshimatsu2017superconductivity} Magn\'eli phase. Studies of mixed eutectic phases indicated a correlation between the coexistence of c-$\mathrm{TiO_{1+\delta}}$, cr-$\mathrm{Ti_2O_3}$ and $\mathrm{\gamma-Ti_3 O_5}$ phases with enhanced T$_C$  hinting at the non-trivial role of interactions at interfaces between Ti$_x$O$_y$ phases in enhancing T$_C$.\cite{kurokawa2017effects} Thus, a critical step in understanding the enhanced transition temperature, $\mathrm{T_c}$, in Ti$_x$O$_y$ thin films involves elucidating the atomic-scale structure and composition of the system as a function of growth conditions (growth temperature, oxygen pressure, film thickness) using synthesis techniques which allow for the control of the film stoichiometry.

\begin{figure}[h]
\centering
\includegraphics[width=1\textwidth]{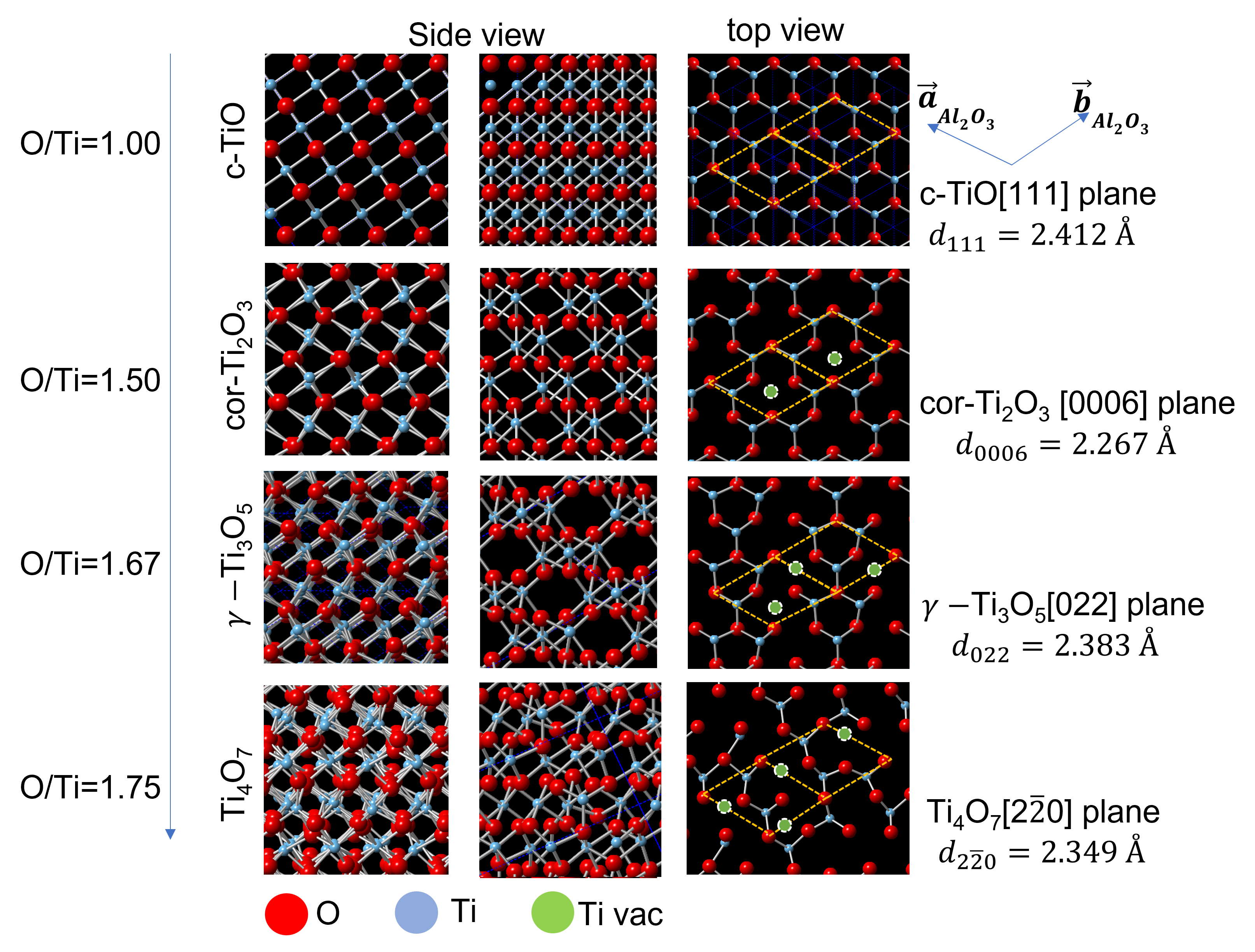}
\caption{Atomic structure  of cubic c-TiO, corundum $\mathrm{cr-Ti_2O_3}$,  $\mathrm{\gamma-Ti_3O_5}$ and $\mathrm{Ti_4O_7}$ along projections in the (a) [10$\bar{1}$0] (b) [11$\bar{2}$0]  and (c) [0001] $\mathrm{\alpha-Al_2O_3}$ directions. }
\label{fig:epitaxy}
\end{figure}

The relation between the transport properties and structural phases of $\mathrm{Ti_xO_y}$ was previously investigated for thin films grown by pulsed laser deposition (PLD) and molecular beam epitaxy (MBE) on (0001)-oriented $\mathrm{\alpha-Al_2O_3}$.\cite{yoshimatsu2017superconductivity, yoshimatsu2018large, li2018orthorhombic, fan2018quantum, zhang2017quasi, kurokawa2017effects, zhang2017enhanced, wang2017enhanced} Figure \ref{fig:epitaxy} shows the lattice structures of the lattice planes of $\mathrm{Ti_xO_y}$ phases grown epitaxially on (0001)-oriented $\mathrm{\alpha-Al_2O_3}$. A summary of reported phases and their superconducting T$_C$s is given in Table \ref{table:Survey}. Li \textit{et. al.} reported an orthorhombic o-$\mathrm{Ti_2 O_3}$ phase for films grown by PLD above 650$^{\circ}$C as evidenced by high-resolution transmission electron microscopy (TEM), Raman spectroscopy and specular X-ray diffraction measurements (XRD).\cite{li2018observation,li2018orthorhombic} They observed a dependence of $\mathrm{T_c}$ on the film thickness of the $\mathrm{o-Ti_2 O_3}$ phase with a maximum $\mathrm{T_c}$ of 8 K for 168 nm thick films.\cite{li2018observation} Low temperature growth (below 650 $^o$C) led to the stabilization of the insulating bulk trigonal $\mathrm{cr-Ti_2 O_3}$ phase. Fan \textit{et. al.} investigated the growth oxygen pressure dependence between 4.5$\times 10^{-6}$ to 6.7$\times 10^{-6}$ Torr for 80 nm thick films and found a suppression of superconductivity at high oxygen growth pressures.\cite{fan2018quantum}  The conclusion of a c-TiO structure was based on the analysis of the film layers close to the substrate by TEM and XRD measurements.\cite{zhang2017enhanced} It is important to note that while multiple structures have been proposed/observed, the T$_C$s  are consistent with the reported thickness-dependence.\cite{zhang2019quantum, li2018observation} Additionally, local TEM measurements indicated a transitional c-TiO interface layer for  o-$\mathrm{Ti_2O_3}$.\cite{li2018observation} A study of eutectic $\mathrm{Ti_xO_y}$ prepared by varying the growth and post-growth Ar annealing conditions showed a variation of T$_C$ with the relative fractions of c-TiO, $\mathrm{\gamma-Ti_3O_5}$ and $\mathrm{cr-Ti_2O_3}$. Thus, the complexity of the system requires a systematic investigation of how specific growth conditions influence the relative fractions of the phases of $\mathrm{Ti_xO_y}$ and the resulting effect on the transport properties.

\begin{table}[]
    \centering
  %  \begin{tabular}{|c|c|c|c|c|c{1cm}|}
     \begin{tabular}{|p{2cm}|p{2cm}|p{2cm}|p{2cm}|}
     \hline
          \textbf{Substrate (Method)} &	\textbf{Thickness}  &	\textbf{Structure}	& $\mathbf{T_c}$  \\
              \hline
         \hline 
         $\mathrm{Al_2 O_3}$ (PLD) & 80nm & $\mathrm{TiO_{1\pm\delta}}$ & 7.4 K  \cite{zhang2017enhanced}  \\
     \hline
          (Sintering) & bulk & $\mathrm{TiO_{1\pm\delta}}$   & 5.5 K  \cite{wang2017enhanced}  \\
     \hline
       $\mathrm{Al_2 O_3}$(PLD) & 168 nm & o-$\mathrm{Ti_2 O_3}$  &8 K  \cite{li2018observation}\\ 
     \hline
            $\mathrm{Al_2 O_3}$(PLD) & 120 nm  & o-$\mathrm{Ti_2 O_3}$   &I (no T$_C$)  \cite{li2018orthorhombic} \\
            
   \hline
            $\mathrm{Al_2 O_3}$ (PLD ) & 120 nm &  $\mathrm{\gamma-Ti_3 O_5}$ & 7.1  \cite{yoshimatsu2017superconductivity}  \\ 
            
   \hline
            LSAT, $\mathrm{Mg Al_2 O_4}$ (PLD ) & 120 nm & $\mathrm{Ti_4 O_7}$ &3.0   \cite{yoshimatsu2017superconductivity}   \\

   \hline
           $\mathrm{Al_2 O_3}$ (PLD ) & 80 nm & $\mathrm{TiO_{1\pm\delta}}$ & 1.4-6 K   \cite{fan2018quantum}    \\ 
            
   \hline
           $\mathrm{Al_2 O_3}$ (PLD) & 150 nm & cr-$\mathrm{Ti_2O_3}$ & I(no $T_C$)   \cite{yoshimatsu2018large}    \\ 
           
        \hline
          
    \end{tabular}
    \caption{Summary of reported $\mathrm{Ti_xO_y}$ structures and superconducting transition temperatures, $\mathrm{T_c}$.}
    \label{table:Survey}
\end{table}
In this letter, we investigate the transport and structural properties of $\mathrm{Ti_xO_y}$ films grown by MBE to elucidate the effect of tuning the film stoichiometry and thickness on the superconducting properties of $\mathrm{Ti_xO_y}$ thin films. Here, we perform high-resolution synchrotron X-ray diffraction measurements on $\mathrm{Ti_xO_y}$ films grown on (0001)-oriented $\alpha-\mathrm{Al_2O_3}$ at 850 $^o$C. The film stoichiometry is tuned by controlling the growth oxygen pressure, $P_{Ox}$, Ti flux rate, and film thickness. $P_{Ox}$ is varied from $4 \times 10^{-8}$ Torr to $1 \times 10^{-6}$ Torr. Films grown at low oxygen partial pressures ($P_{Ox} \leq 1 \times 10^{-7}$ Torr) are found to be metallic with carrier concentrations ranging from 3$\times 10^{22}$ cm$^{-3}$ at 300 K to 6$\times 10^{21}$ cm$^{-3}$ at 10 K. Metallicity is correlated with the formation of oxygen-deficient c-$\mathrm{TiO_{1-\delta}}$. By tuning the $P_{Ox}$, the Ti flux rate and the film thickness, an insulating-superconducting transition is observed in 85 nm thick films which are characterized by a thin c-$\mathrm{TiO_{1\pm\delta}}$ interfacial buffer layer and the formation of the $\mathrm{\gamma-Ti_3O_5}$ phase. Strain relaxation is observed in $\mathrm{\gamma-Ti_3O_5}$ with increasing film thickness and correlates with a thickness-tuned T$_C$ observed in Ti$_x$O$_y$ films.\cite{zhang2019quantum, li2018observation} %  By tuning the $P_{Ox}$ and the Ti flux rate a phase mixture of c-TiO, $\mathrm{cr-Ti_2O_3}$, $\mathrm{Ti_4O_7}$, and $\mathrm{\gamma-Ti_3O_5}$ is formed. The distribution of the phases are mapped using synchrotron X-ray microdiffraction.
%The structural properties of the films are investigated by atomic force microscopy (AFM), x-ray diffraction (XRD), and x-ray reflectivity (XRR). 
%For $\mathrm{Ti_xO_y}$ thin films, a wide range of structural phases and lattice parameters have been reported and are summarized in Table \ref{table:Survey}. %\note{The $\mathrm{Ti_3O_5}$ phase is ruled out by Li et al through their XPS measurements} $\mathrm{Ti_3 O_5}$ phase ruled out by Li et al. \cite{li2018observation} by XPS

\begin{figure*}[ht]
\centering
\includegraphics[width=1\textwidth]{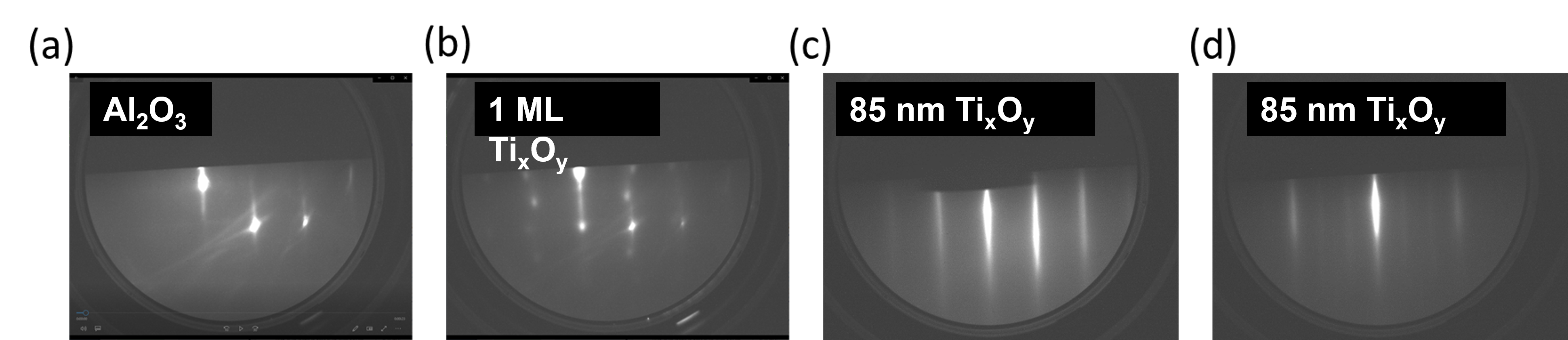}
\caption{Evolution of the surface RHEED pattern for the growth of 85 nm thick $\mathrm{Ti_xO_y}$ film on (0001)-oriented $\alpha-\mathrm{Al_2O_3}$ at P$_{O2}=2\times 10^{-7}$ Torr and a growth rate of 2.5 \AA{}/min (Sample E). (a) Diffraction pattern of the initial $\mathrm{Al_2O_3}$ surface (b) after 1 ML $\mathrm{Ti_xO_y}$ along the [11$\bar{2}0$] azimuth, (c) after 85 nm along the [11$\bar{2}0$] and (d)[10$\bar{1}0$] directions.}
\label{fig:RHEED}
\end{figure*}

\section{Results and Discussion}

%\subsection{Growth}
 $\mathrm{Ti_xO_y}$ films with thicknesses 25-85 nm were grown by oxide MBE. Prior to growth, the $\mathrm{Al_2O_3}$ substrates were annealed at 1100 $^o$C in a tube furnace tube for 12 hours. Atomic force microscope images show smooth surfaces with atomic steps. The films were grown by deposition of Ti from an effusion cell under partial pressures of molecular oxygen  ranging from 4$\times10^{-8}$ to 1$\times10^{-6}$ Torr at a substrate temperature of 850 $^o$C. The growth rates were determined from the Ti fluxes measured by a quartz crystal monitor and X-ray reflectivity measurements to be 1.3-2.5 \AA{}/min. The sample growth conditions are summarized in Table \ref{tab:samples}. After growth, the films were cooled down at a rate of 25 $^o$C/min from the deposition temperature to room temperature in vacuum.

\begin{table*}[ht]
    \centering
    \begin{tabular}{|c|c|m{3cm}|c|m{3cm}|m{3.5cm}|c|}
     \hline
         	Sample & Thickness  & Ti flux rate ($atoms/cm^2s$) &	$\mathrm{P_{Ox} (Torr)}$& Normalized $P_{Ox}/Ti_{rate}$ & Phase & Transport \\
         \hline 
         \hline
         A  &   25 nm & 1.2 x10$^{13}$  & $3  \times 10^{-7}$&2.9 & $\mathrm{cr-Ti_2O_3}$ & I \\
         \hline 
        B  &   25 nm & 1.2 x10$^{13}$  & $4  \times 10^{-8}$ &0.4 & c-TiO$_{1-\delta}$ , $\gamma-Ti_3O_5 $& M \\
         \hline 
         C  &   25 nm & 1.2 x10$^{13}$  & $8  \times 10^{-8}$ &0.8 & c-TiO$_{1-\delta}$ , $\gamma-Ti_3O_5 $ & M\\
        \hline
         D  &   45 nm & 2.4 x10$^{13}$  & $2 \times 10^{-7}$ &1.0 & c-$\mathrm{TiO_{1-\delta}, \gamma-Ti_3O_5}$ & M,SC\\
        \hline
         E  &   85 nm & 2.4 x10$^{13}$ & $2 \times 10^{-7}$ &1.0 & c-$\mathrm{TiO, \gamma-Ti_3O_5}$ & I,SC\\
        \hline
        F  &   80 nm & 1.7  x10$^{13}$ & $2  \times 10^{-7}$ &1.4 & c-$\mathrm{TiO/\gamma-Ti_3O_5}$,\newline ${cr-Ti_2O_3, Ti_4O_7}$ & I, S\\
        \hline

    \end{tabular}
    \caption{Summary of growth conditions, film thickness and measured phases and transport properties of MBE-grown $\mathrm{Ti_xO_y}$ films on (0001)-oriented $\alpha-Al_2O_3.$ I, M and SC refer to the insulating, metallic and superconducting phases, respectively. The $P_{Ox}/Ti_{rate}$ are normalized to the values for the superconducting Sample E. }
    \label{tab:samples}
\end{table*}

Figure \ref{fig:RHEED}(a) shows an \textit{in-situ} reflection high energy diffraction (RHEED) image of the initial $\mathrm{Al_2O_3}$ substrate surface prior to deposition of the $\mathrm{Ti_xO_y}$ films at $P_{Ox}=2\times 10^{-7}$ Torr. Figure \ref{fig:RHEED}(b) shows the RHEED pattern after the deposition of 1 ML of $\mathrm{Ti_xO_y}$. Roughening and relaxation are evident from the spotty nature of the RHEED pattern and the development of streaks with a closer spacing than the diffraction from the substrate. A streaky 2D pattern emerges after the 2nd ML and remains for the entire film growth. Figure \ref{fig:RHEED}(c) and \ref{fig:RHEED}(d) show the RHEED patterns after the growth of an 85 nm thick film along the [11$\bar{2}0$] and [10$\bar{1}0$] directions respectively. The narrow streaks are indicative of a smooth 2D surface. Along the [10$\bar{1}0$] direction, a 3x reconstruction is observed.A comparasion of the final RHEED patterns of Samples A-F is shown in Figure S1 of the supplemental materials.\cite{suppl} Atomic force microscope images of the as-grown films(See Figure S2 of supplemental materials)\cite{suppl} show atomically flat layers with step-like features identical to the substrate.

\subsection{Transport Properties}

\begin{figure*}[pt]
\centering
\includegraphics[width=0.8\textwidth]{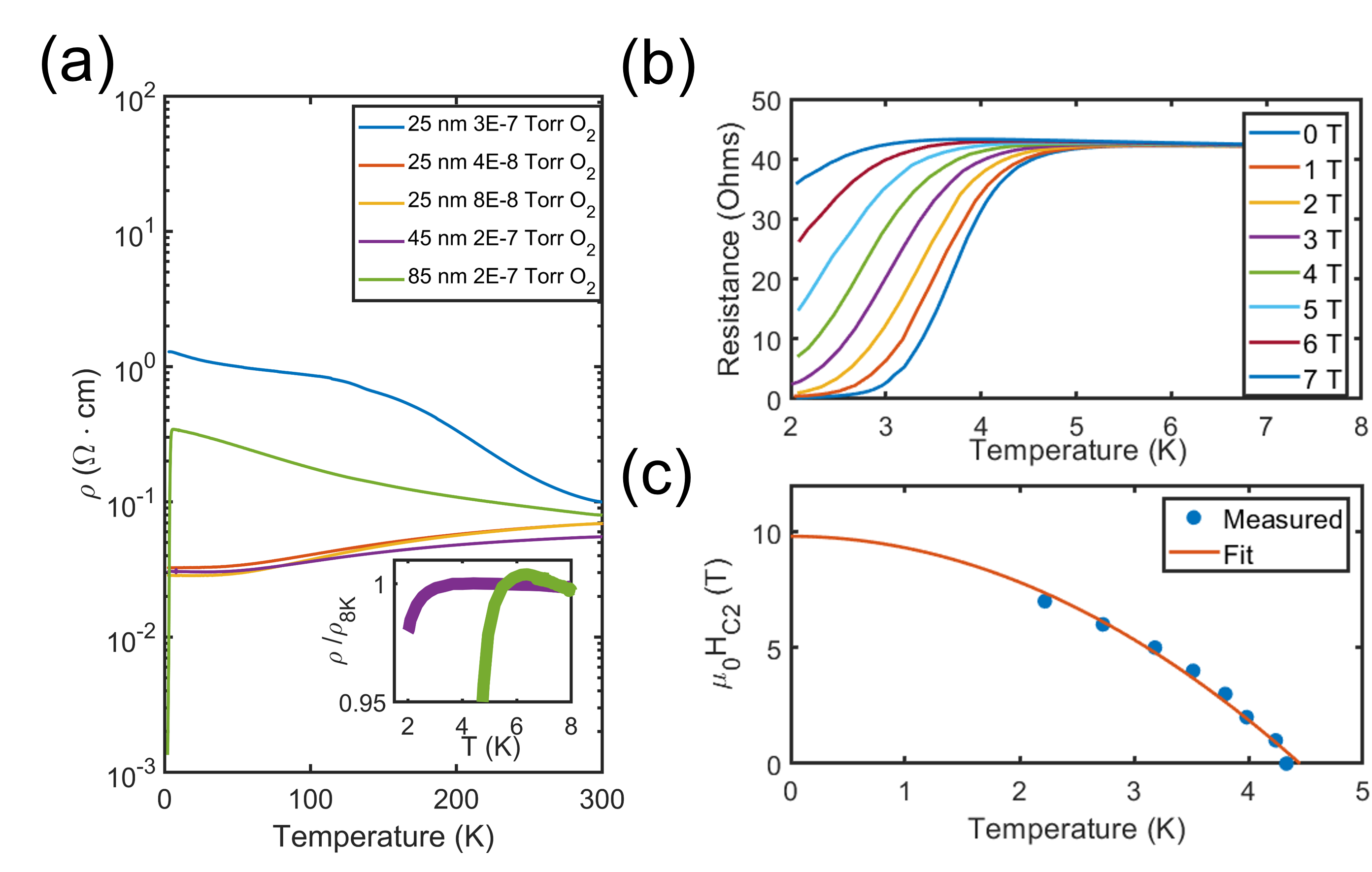}
\caption{ (a) Comparison of sheet resistance for MBE-grown $\mathrm{Ti_xO_y}$ films as a function of growth oxygen pressure and film thickness. The inset shows the drop in resistivity at low temperatures for the superconducting samples (Samples D and E). (b) Resistivity as a function of magnetic field applied in-plane for 85 nm thick $\mathrm{Ti_xO_y}$ film (Sample E) grown at $P_{Ox}=2\times10^{-7}$ Torr. (c) Relationship between critical magnetic field and T$_C$ and the corresponding Werthamer-Helfand-Hohenberg (WHH) fit. }
\label{fig:transport}
\end{figure*}

The dependence of the transport properties on the film thickness and growth conditions is determined by comparing the resistivities of a series of $\mathrm{Ti_xO_y}$ films with thicknesses between 25 nm and 85 nm for $4\times 10^{-8} \leq P_{Ox}\leq 3\times 10^{-7}$  Torr in Figure \ref{fig:transport}(a). The transport measurements are performed in the Van der Pauw configuration using Au contacts deposited on the corners of 5mm $\times$ 5 mm samples. Films grown with $P_{Ox} \leq 1\times10^{-7}$  Torr are metallic for thicknesses less than 45 nm. Growth at higher oxygen pressures ($P_{Ox} \geq 3\times10^{-7}$ Torr) results in insulating  transport properties. 

At the intermediate growth pressure of $P_{Ox} = 2\times10^{-7}$ Torr, the transport properties depend on the film thickness and the Ti flux rate. The 45 nm film grown at $P_{Ox} = 2\times10^{-7}$ Torr (Sample D) is metallic and undergoes a metal-superconducting transition at $\sim$ 3.7 K. The thicker 85 nm film (Sample E) is insulating below 300 K and transitions to a superconducting state at 4.5 K. The absence of superconductivity above 2 K for the 25 nm thick films, for the Ti flux rates investigated, is consistent with the thickness dependence of T$_C$ observed for PLD grown films.\cite{li2018observation} Films grown with $P_{Ox}>8\times 10^{-7}$ Torr are insulating due to the formation of TiO$_2$ as evidenced by X-ray diffraction measurements.

%Hall measurements are performed between 300 K and 5 K and the measured mobilities, $\mu$, and 3D carrier concentrations, $n$, are summarized in Figure \ref{fig:transport}(b) and \ref{fig:transport}(c). For the 25 nm metallic films grown at $4\times 10^{-8}$ and $8\times 10^{-8}$ Torr, $\mu$ is 0.7 $cm^2V^{-1}s^{-1}$ at 300 K and increases to ~7.5 $cm^2V^{-1}s^{-1}$ at 3 K while $n$ decreases from $3\times 10^{22}$ cm$^{-3}$ at 300 K to $6\times 10^{21}$ cm$^{-3}$  at 5 K. The $n$ are slightly larger than the values reported for nonstoichiometric $TiO_x$ ($x>1$) films.\cite{fan2018quantum, akazawa2011transparent, yoshimatsu2020metallic} For the 25 nm insulating sample grown at $3\times 10^{-7}$ Torr, $\mu=27$ $cm^2V^{-1}s^{-1}$ between 300 K and 5 K and $n=5\times 10^{20}$ cm$^{-3}$ at 300 K and  decreases to $4\times 10^{19}$ cm$^{-3}$ at 5 K.

%For the superconducting 85 nm sample, $\mu$ reduced between 300 K and 5 K from 0.4 to 0.1 $cm^2V^{-1}s^{-1}$ while the $n$ has a temperature-independent value to $4\times 10^{22}$ cm$^{-3}$.% \textbf{what is expected for superconductors?}

Superconductivity in the 85 nm film (Sample E) is confirmed by measuring the resistance as a function of a magnetic field, $H$ applied parallel to the sample surface. Figure \ref{fig:transport}(b) shows the suppression of T$_C$ for $0 T <H< 7 T$. The upper critical field $H_{C2}(T)$ is defined by a 90$\%$ drop in the normal state resistance. Figure \ref{fig:transport}(c) shows a plot of $H_{C2}$ as a function of temperature. The results are fit to the Werthamer-Helfand-Hohenberg (WHH) equation $H_{C2}(T)=H_{C2}(0)[1-(\frac{T}{T})^2]$.\cite{werthamer1966temperature} From the fit, the upper critical field at 0 K , $H_c2(0)$ is determined to be 9.8 T. The coherence length, $\xi$, is determined to be 5.7 nm from the the Ginzburg-Landau superconducting coherence length relation, $\xi =[(\hbar/2e)/( H_{c2}(0))]^{1/2}$. The measured coherence length is comparable to previous reports. \cite{zhang2017enhanced, fan2018quantum}% \textbf{T$_C$ is determined from the max of $\frac{dR}{dT}$. check for more formal definition of $\mathrm{T_c}$}.

The insulating-superconducting transition has been previously observed in Ti$_x$O$_y$ films\cite{zhang2019quantum, hulm1972superconductivity} and another 3D superconductor, BaPb$_{1-x}$Bi$_x$O$_3$  which also exhibits a strong dependence of T$_C$ on the film thickness and the prescence of multiple polymorphs (tetragonal and orthorhombic fractions of BaPb$_{1-x}$Bi$_x$O$_3$).\cite{harris2018superconductivity} Thus, structural disorder arising from local fluctuations in oxygen content may enhance disorder leading to the insulator-superconductor transition.

\subsection{XRD Structure}
\begin{figure}[ht]
\centering
\includegraphics[width=1\textwidth]{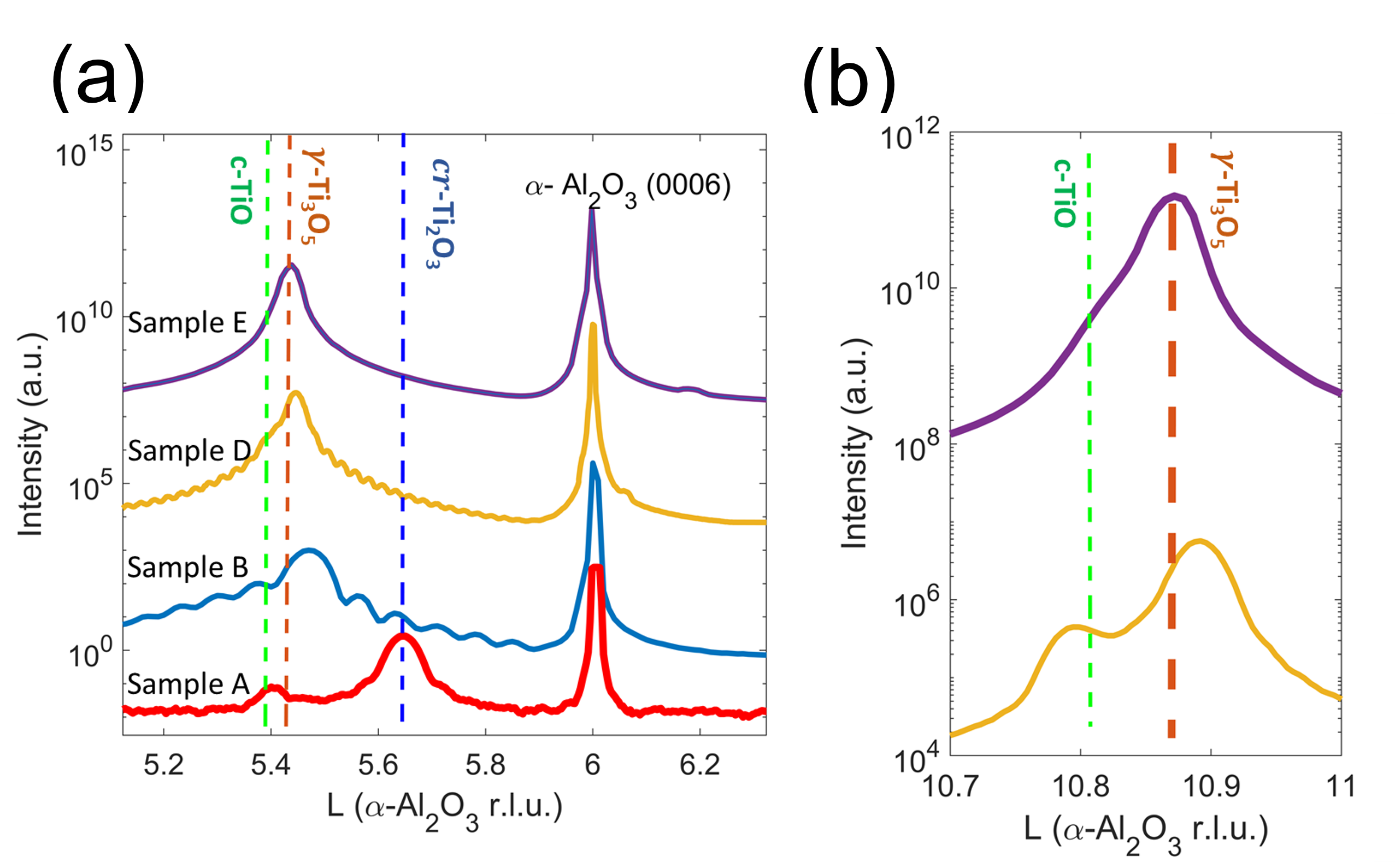}
\caption{(a) Comparison of specular X-ray diffraction around the $\mathrm{\alpha-Al_2O_3}$ (0006) Bragg peak for a Samples A (25 nm, $P_{Ox}=3\times10^{-7}$ Torr),  B (25 nm, $P_{Ox}=4\times10^{-8}$ Torr), D( 45 nm, $P_{Ox}=2\times10^{-7}$ Torr) and E( 85 nm, $P_{Ox}=2\times10^{-7}$ Torr) Ti$_x$O$_y$ films grown by MBE. (b) Corresponding reflections for Sample D and E at higher q.   }
\label{fig:specular}
\end{figure}

The relationship between the synthesis conditions, crystal structure, and the transport properties is determined by synchrotron XRD measurements at the 33ID beamline at the Advanced Photon Source. Figure \ref{fig:specular}(a) shows a specular scan around the substrate (0006) Bragg peak measured with an X-ray energy of 16 KeV ($\lambda=0.774$ \AA{}) for a series of $\mathrm{Ti_xO_y}$ samples. For Sample A grown with the highest $P_{Ox}/Ti_{rate}$ ratio, the main Bragg peak corresponds to the cr-$\mathrm{Ti_2O_3}$ phase. A second peak is observed corresponding to a slightly O-rich c-$\mathrm{TiO}$ layer.

The diffraction intensities for the 25 nm metallic $\mathrm{Ti_xO_y}$ film (Sample B) comprise of a broad shoulder with lattice spacing d=2.423 \AA{} and a main peak with d=2.375 \AA{}. The shoulder and main peaks correspond, respectively, to the (111) Bragg reflection of oxygen-deficient c-$\mathrm{TiO_{-\delta}}$ and the (022) Bragg peak of  $\mathrm{\gamma-Ti_3O_5}$ compressively strained to the buffer layer.  

While weak insulating behavior has been reported for c-$\mathrm{TiO_{1+\delta}}$, the observation of metallicity in Samples B and C arises from the stabilization of Ti-rich/oxygen poor c-$\mathrm{TiO_{1-\delta}}$ due to the low growth oxygen pressures. For $0.05\leq \delta \leq 0.2$, Hulm \textit{et. al.} show metallicity for as-cast single-phase c-$\mathrm{TiO_{1-\delta}}$ samples.\cite{hulm1972superconductivity} An expansion in the lattice parameter is expected for the oxygen deficient c-$\mathrm{TiO_{1-\delta}}$,\cite{hulm1972superconductivity} hence, from fits to the 00L data, the broad shoulder at low \textit{L} ($c_{measured}$=4.19 \AA{}) is assigned to a $\sim$15 \AA{} thick metallic c-$\mathrm{TiO_{1-\delta}}$ layer which dominates the resistivity measurements in Figure \ref{fig:transport}(a). Annealing Sample B in flowing O$_2$ for 3 hours at 600 $^o$C leads to the oxidation of both layers and a shift in the Bragg peaks to d=2.324 \AA{} corresponding to the formation of cr-$\mathrm{Ti_2O_3}$ and a metal-insulator transition.

The 45 nm metallic Sample D also has a main peak at d=2.385 \AA{} corresponding to the (022) peak of $\gamma -Ti_3O_5$ and a shoulder with d=2.4171 \AA{} corresponding to slightly oxygen-deficient c-$\mathrm{TiO_{1-\delta}}$. The metallicity observed for this sample is again, due to the metallic oxygen-deficient c-$TiO_{1-\delta}$ layer. Around 3.7 K, a decrease in the resistivity is observed (Figure \ref{fig:transport}(a) inset) due to the superconducting transition in the $\gamma -Ti_3O_5$ layer.\cite{yoshimatsu2017superconductivity} 

The superconducting Sample E has a main Bragg peak at d=2.388 \AA{} corresponding to the (022) peak of relaxed $\gamma -Ti_3O_5$. Close to the (044) $\gamma -Ti_3O_5$ in Figure \ref{fig:specular}(b), we observe a shoulder at lower q which corresponds to the (222) reflection of cubic stoichiometric TiO ($c_{measured}=4.167$ \AA{}). The metallic phase is strongly suppressed, possibly, due to the longer deposition time which allows for complete oxidation of the $c-TiO$ buffer layer. The transport properties of the Sample E in Figure \ref{fig:transport}(a) are dominated by the  superconducting relaxed $\gamma -Ti_3O_5$ phase.\cite{fan2019structure}

\begin{figure*}[ht]
\centering
\includegraphics[width=1\textwidth]{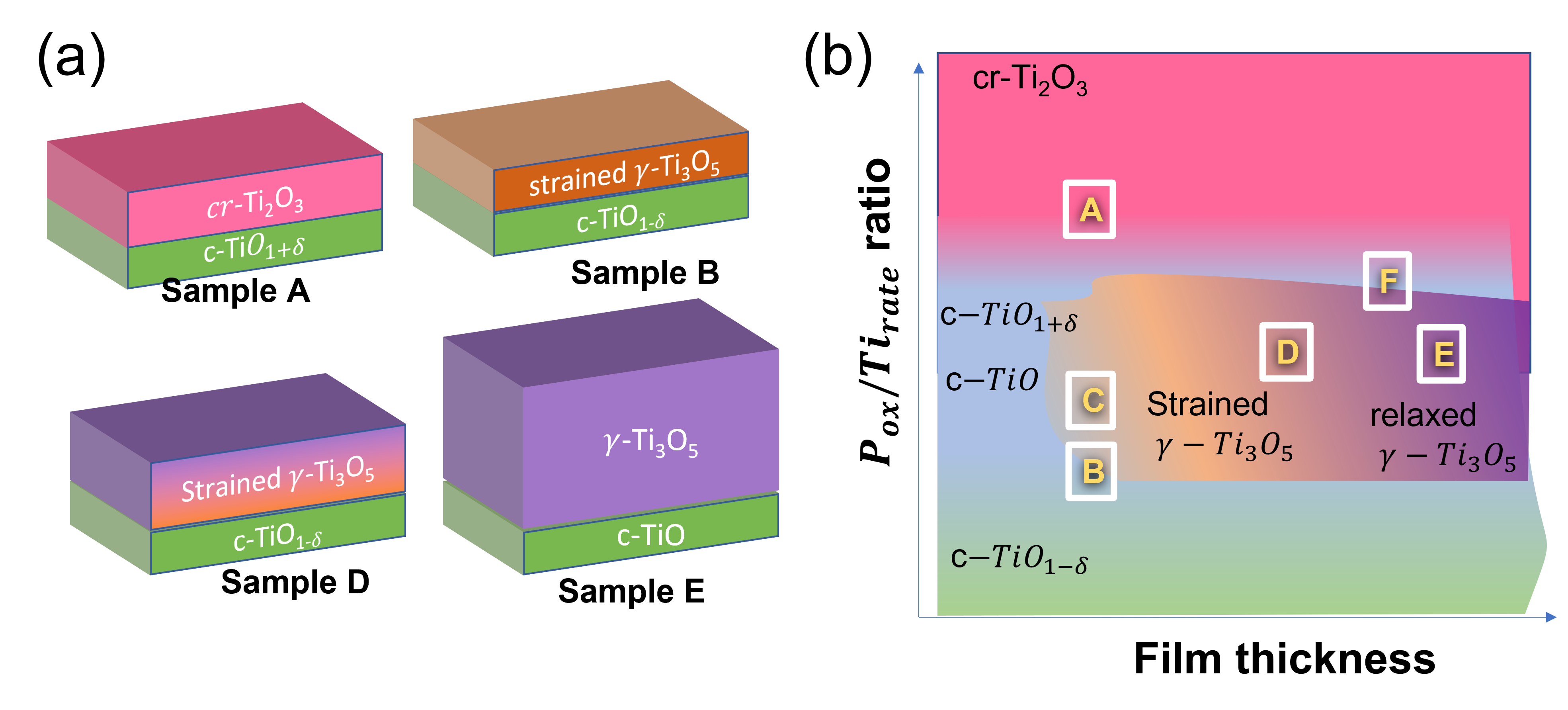}
\caption{(a) Schematic structures of samples A,B,D and E determined from X-ray diffraction measurements and (b) the corresponding phase diagram showing the relationship between the ratio of the growth oxygen pressure $P_{ox}$ to the Ti flux rate and the film thickness.}
\label{fig:phase}
\end{figure*}

Based on the X-ray diffraction measurements, the emergent picture of the structural profiles of the films under different growth conditions is summarized in Figure \ref{fig:phase}. Oxygen-poor growth of $\mathrm{Ti_xO_y}$ on (0001)-oriented $\mathrm{\alpha-Al_2O_3}$ at high growth temperatures leads to the formation of a thin (111)-oriented cubic $\mathrm{TiO_{1-\delta}}$ metallic buffer layer. As the film thickness increases, if the oxygen pressure is sufficiently high, the structure transforms into the Magn\'eli $\mathrm{\gamma-Ti_3O_5}$ phase. The $\mathrm{\gamma-Ti_3O_5}$ is initially strained to the buffer layer and the strain is relaxed as the film thickness is increased. Since pressure is known to suppress superconductivity in $\mathrm{Ti_xO_y}$,\cite{zhang2017enhanced} this picture is consistent with the observed increase in T$_C$ with increasing film thickness.\cite{fan2019structure, li2018observation}

\subsection{Phase segregation in $\mathrm{Ti_xO_y}$}
The transport and structural results indicate a strong correlation between the growth conditions and film thickness on the physical and structural properties of the $\mathrm{Ti_xO_y}$ films. Stoichiometric c-TiO has an NaCl structure with lattice constant c=4.177 \AA{}. The NaCl structure is stable for oxygen concentrations ranging from 0.8 to 1.3 with T$_C$ increasing from 0.5 K to 1 K for bulk c-$TiO_x$.\cite{hulm1972superconductivity, reed1972superconductivity, banus1972electrical, li2021single} The tendency of $\mathrm{Ti_xO_y}$ to form Magn\'eli phases and related polymorphs of the form $\mathrm{Ti_nO_{2n-1}}$ suggests that the oxygen stoichiometry and the growth conditions strongly influence the structure and physical properties of the thin films. %Recent results indicate that the interfaces in $\mathrm{Ti_xO_y}$ eutectics and at the interfaces between oxygen-rich and oxygen-poor $\mathrm{Ti_xO_y}$ phases may play a role in the enhanced T$_C$s  observed in $\mathrm{Ti_xO_y}$ thin films.\cite{yoshimatsu2017superconductivity}  % The enhanced T$_C$ in $\mathrm{Ti_xO_y}$ thin films and nanostructures has been attributed to enhanced conduction electron-substrate phonon interaction and/or strain effects.\cite{zhang2017enhanced}.   $\mathrm{Ti_4O_7}$(n=4) and $\gamma$-$\mathrm{Ti_3O_5}$ (n=5) films have T$_C$s  of 3.0 and 7.1 K, respectively.

\begin{comment}
\begin{figure}[h]
\centering
\includegraphics[width=0.5\textwidth]{images/Compare_B006_B007.png}
\caption{Comparison of specular diffraction. Insert shows the magnified (222) film peak. }
\label{fig:00L}
\end{figure}

\end{comment}

\begin{figure*}[ht]
\centering
\includegraphics[width=1\textwidth]{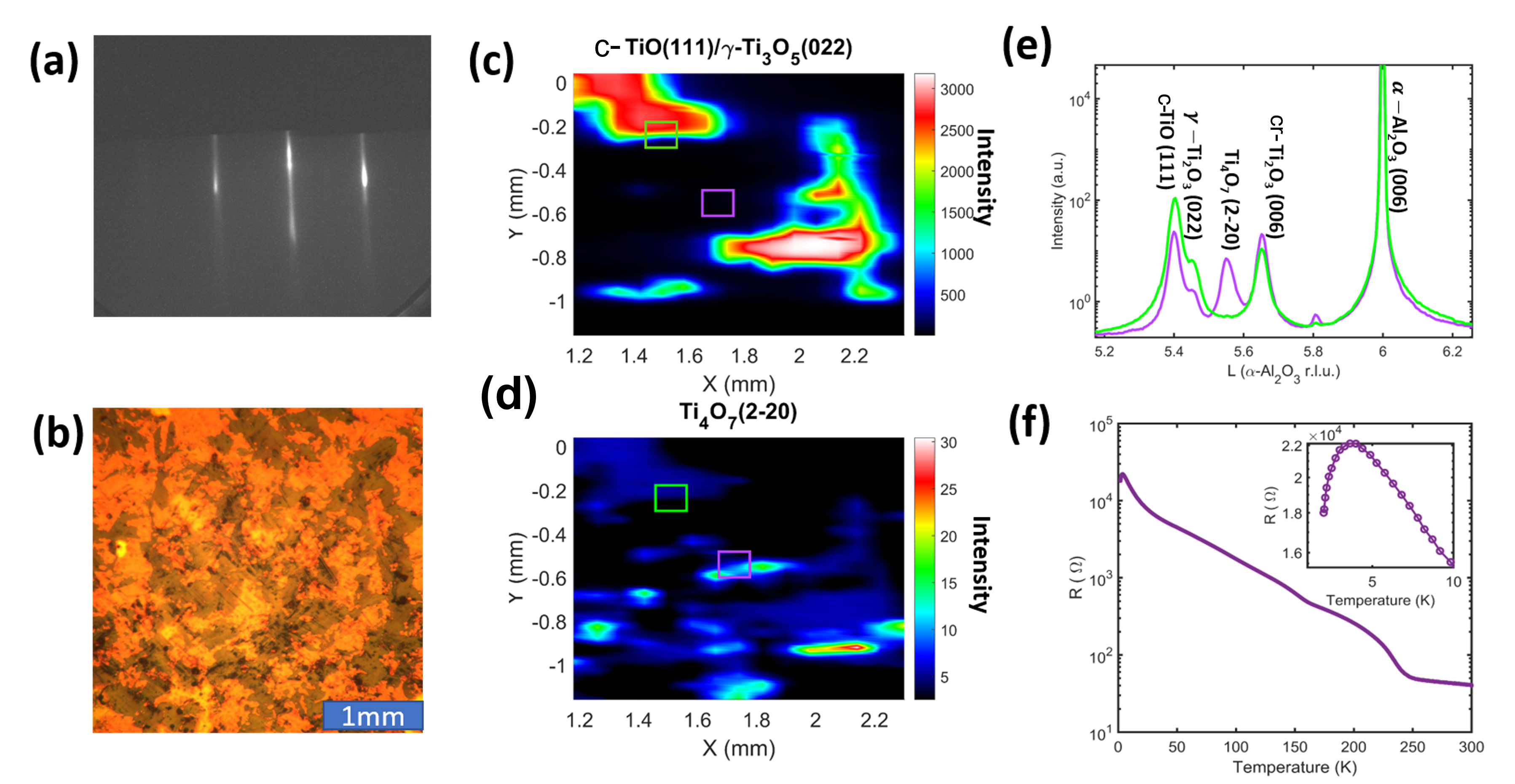}
\caption{Structural and transport properties of a 80  nm thick $\mathrm{Ti_xO_y}$ film grown at Ti flux rate of 1.8 \AA{}/min in P$_{O2}=2\times10^{-7}$ Torr (Sample F). (a) Reflection high energy diffraction for 80 nm $\mathrm{Ti_xO_y}$ on 001-oriented Al$_2$O$_3$. (b) Optical micrograph of film indicating structural domains. High-resolution synchrotron X-ray diffraction raster maps around Bragg peaks observed at diffraction conditions fixed for (c) (111) c-TiO Bragg peak and (d) $\mathrm{Ti_4O_7}$ Bragg peak indicating structural phase separation. (e) Specular diffraction along the substrate 00L direction at different points on the sample.  (f) Resistance versus temperature. The inset shows the superconducting transition at low temperature. }
\label{fig:B012}
\end{figure*}

To further investigate the delicate balance between film stoichiometry and the transport properties of the $\mathrm{Ti_xO_y}$ system, we consider the structural properties of an eutectic film formed by growing at a reduced Ti rate with the oxygen pressure of 2$\times 10^{-7}$ Torr. The slight increase in the O/Ti ratio is expected to allow for the thermodynamic stabilization of oxygen-rich $\mathrm{Ti_xO_y}$ phases such as $\mathrm{cr-Ti_2O_3}$. 

The initial RHEED images for the sample grown at the lower Ti rate shows diffraction spots indicative of surface roughening, however, the final RHEED image in Figure \ref{fig:B012}(a) shows well defined 2D streaks after the growth of 80 nm. Figure \ref{fig:B012}(b) shows an optical image of the as-grown film. Clear ‘dark’ and ‘bright’ regions are observed indicating macroscopic phase separation. These variations are not observed for the uniform films previously discussed (Figure S3 of supplemental materials)\cite{suppl}. To determine the differences in the structure of the observed phases, we perform micro-diffraction experiments at the Advanced Photon source where the X-ray beam is focused to a 20$\mu$m x40$\mu$m spot size using a Kirkpatrick–Baez mirror. Figure \ref{fig:B012}(c) and \ref{fig:B012}(d) shows diffraction intensity maps with the diffraction conditions fixed at the c-TiO (111) Bragg reflection and the $\mathrm{Ti_4O_7}$ (2-20) reflection. The maps show that the two phases are not uniformly distributed throughout the sample. Figure \ref{fig:B012}(e) shows (00L) scans at locations close to the c-TiO phase and the $\mathrm{Ti_4O_7}$ domain. The scans show that the intensity of the $\mathrm{Ti_4O_7}$ is suppressed in regions of the film with increased fractions of the c-TiO and $\mathrm{\gamma-Ti_3O_5}$ phase. At both locations surveyed in  Figure \ref{fig:B012}(e), a cr-$\mathrm{Ti_2O_3}$ peak is present indicating that the c-TiO/$\mathrm{\gamma-Ti_3O_5}$ and $\mathrm{Ti_4O_7}$ grains are located in a matrix of the insulating $\mathrm{cr-Ti_2O_3}$ phase (The regions with 0 intensity in Figure \ref{fig:B012}(c) and (d)).
 The $\mathrm{Ti_4O_7}$ grains have lateral dimensions on the order of 100 $\mu m$ while the c-TiO/$\mathrm{\gamma-Ti_3O_5}$ grains are an order of magnitude larger in dimensions. 

The transport properties of the eutectic sample is shown in Figure \ref{fig:B012}(f). The multiple electronic transitions are consistent with the coexistence of multiple structural phases. Transitions in the resistivity are observed at ~250 K and 150 K which are consistent with reported transitions for cr-$\mathrm{Ti_2O_3}$ \cite{kurokawa2017effects} and $\mathrm{Ti_4O_7}$ \cite{yoshimatsu2017superconductivity}. The resistivity drops at ~3 K is indicative of a superconducting transition expected for either $\mathrm{Ti_4O_7}$ or $\mathrm{\gamma-Ti_3O_5}$, however, no zero-resistant state is observed at the instrument’s minimum temperature of 2 K.

\section{Conclusion}
In conclusion, we have used a combination of high-resolution synchrotron X-ray diffraction mapping and temperature-dependent transport to investigate the correlation between film thickness, phase separation and superconductivity in $\mathrm{Ti_xO_y}$ films grown on (0001)-oriented $\mathrm{\alpha-Al_2O_3}$. The films with thicknesses ranging from 25 to 85 nm are grown by MBE where the oxygen stoichiometry of the films is tuned by the oxygen partial pressure and the Ti flux rate during growth. The transport properties are correlated with the P$_{ox}$/Ti$_rate$  ratio and thickness-dependent strain relaxation. A metallic c-$\mathrm{TiO_{1-\delta}}$ buffer layer is formed for films grown under oxygen-poor conditions. A superconducting  Magn\'elli $\mathrm{\gamma-Ti_3O_5}$ layer nucleates on the buffer layer. Strain-relaxation occurs as the film thickness increases and correlates with a thickness-dependent increase in T$_C$. As the P$_{ox}$/Ti flux ratio increases, the buffer composition transitions to an insulating cr-$\mathrm{Ti_2O_3}$ phase. A mixed-phase structure is formed for P$_{ox}$/Ti flux ratios close to the c-TiO/cr-$\mathrm{Ti_2O_3}$  phase boundary.  These results suggest that the thickness-dependence of T$_C$ in $\mathrm{Ti_xO_y}$ is related to a complex interplay between, strain and the nucleation kinetics of $\mathrm{Ti_xO_y}$ phases and polymorphs. Thus, a complete elucidation of phase formation and phase separation and the electronic and structural interactions at inter-phase boundaries will allow for understanding and enhancing T$_C$ in atomically-thin $\mathrm{Ti_xO_y}$ layers.

\section*{Acknowledgments}
The authors acknowledge financial support by the US National Science Foundation under Grant No. NSF DMR-1751455. This work was performed in part at the Analytical Instrumentation Facility (AIF) at North Carolina State University, which is supported by the State of North Carolina and the National Science Foundation (award number ECCS-2025064). This work made use of instrumentation at AIF acquired with support from the National Science Foundation (DMR-1726294). The AIF is a member of the North Carolina Research Triangle Nanotechnology Network (RTNN), a site in the National Nanotechnology Coordinated Infrastructure (NNCI). Use of the Advanced Photon Source was supported by the U.S. Department of Energy, Office of Science, Office of Basic Energy Sciences, under Contract No. DE-AC02-06CH11357.

\section{Data Availability Statement}
The data that support the findings of this study are available from the corresponding author upon reasonable request.

\bibliographystyle{naturemag}

\newpage\newpage
\includepdf[pages={1,{},2-3}]{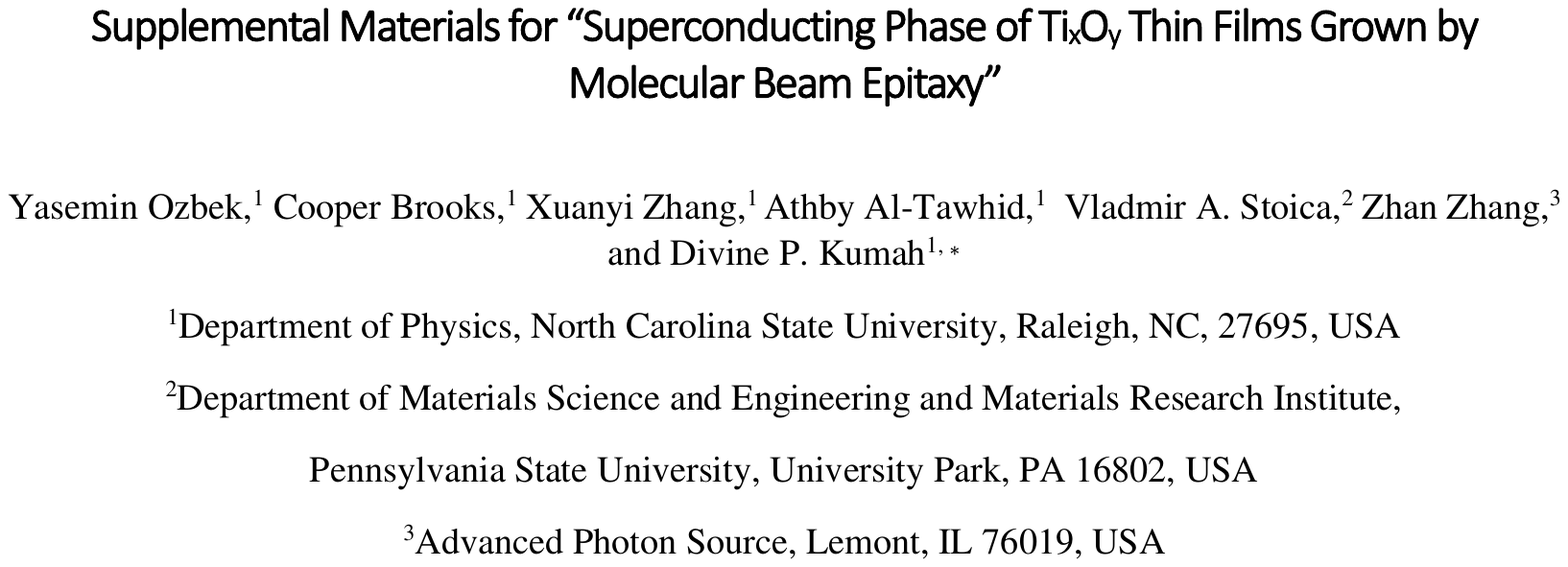}

\end{document}